\ifdefined \singleCol
\documentclass[journal,comsoc,12pt,onecolumn,draftclsnofoot]{IEEEtran}
\else
\documentclass[journal]{IEEEtran}
\fi
\usepackage[T1]{fontenc}
\usepackage[noadjust]{cite}
\usepackage{graphicx}
\graphicspath{{figures/drawings/}{figures/plots/pdf/}}
\usepackage[export]{adjustbox}
\usepackage{subfig}
\usepackage[cmintegrals]{newtxmath}

\usepackage{capt-of}
\usepackage{color,soul}

\begin{document}
	\title{WiSig: A Large-Scale WiFi Signal Dataset for Receiver and Channel Agnostic RF Fingerprinting}
	 

	\author{Samer~Hanna,~\IEEEmembership{Graduate~Student~Member,~IEEE,}
		Samurdhi Karunaratne,~\IEEEmembership{Graduate~Student~Member,~IEEE,}
		and Danijela~Cabric,~\IEEEmembership{Fellow,~IEEE}
		\thanks{The authors are with the Electrical and Computer Engineering Department, University of California, Los Angeles, CA 90095.   	 e-mails: 	\mbox{samerhanna@ucla.edu}, samurdhi@g.ucla.edu , danijela@ee.ucla.edu}
		\thanks{This work was supported in part by the CONIX Research Center, one of six centers in JUMP, a Semiconductor Research Corporation (SRC) program sponsored by DARPA.}
	}
	
	\maketitle

	\begin{abstract}
RF fingerprinting leverages circuit-level variability of transmitters to identify them using signals they send. Signals used for identification are impacted by a wireless channel and receiver circuitry, creating additional impairments that can  confuse transmitter identification.  Eliminating these impairments or just evaluating them, requires data captured over a prolonged period of time, using many spatially separated transmitters and receivers. In this paper, we present WiSig; a large scale WiFi dataset containing 10 million packets captured from 174 off-the-shelf WiFi transmitters and 41 USRP receivers over 4 captures spanning a month.  WiSig is publicly available, not just as raw captures,  but as conveniently pre-processed subsets of limited size, along with the scripts and examples.  A preliminary evaluation performed using WiSig shows that changing receivers, or using signals captured on  a different day can significantly degrade a trained classifier's performance. While capturing data over more days or more receivers limits the degradation, it is not always feasible and novel data-driven approaches are needed.  WiSig provides the data to develop and evaluate these approaches towards channel and receiver agnostic transmitter fingerprinting. 
	\end{abstract}
	
	\begin{IEEEkeywords}
	RF fingerprinting, transmitter identification, WiFi dataset
	\end{IEEEkeywords}

	\IEEEpeerreviewmaketitle

\section{Introduction}

Each transmitter has a unique radio frequency (RF) fingerprint  that is caused by manufacturing variability of its circuit design and components. This variability makes two transmitters, from the same make and model, sending the same waveforms, have slight differences in their  signals. Transmitter (Tx) RF fingerprinting leverages these subtle unintentional differences in the signals to identify transmitters. Tx fingeprinting has the potential of improving the security of wireless networks by verifying the identities of transmitters without imposing any  transmitter-side overheads. 

Since RF fingerprints are not designed but result from manufacturing variability, they need to be extracted using data-driven approaches from the captured signals.  However,  captured signals contain confounding factors besides the transmitter fingerprint: a channel fingerprint is added as the signals propagate through a wireless channel. Also, receivers suffer from manufacturing variability same as transmitters, hence, the receiver RF fingerprint is also embedded in the signal. Data-driven approaches like deep learning are easily confounded by both the channel and receivers. In~\cite{al-shawabka_exposing_2020}, it was shown that evaluating the same transmitter in a different channel can cause the accuracy of a classifier to drop from 85\% to 9\%. Similar trends were reported in~\cite{merchant_toward_2019} for the impact of receivers. Practical deployments of RF fingerprinting systems can involve hundreds of transmitters and need to work over prolonged periods of times where the channel varies. Additionally, aside from micro-scale deployments, many receivers need to participate in the fingerprinting system and training a classifier per receiver might not be practical. These facts drive the need for a channel and receiver agnostic transmitter RF fingerprinting system. The first step in building such a RF fingerprinting system is having the suitable data that enables isolating and evaluating both the channel and receiver impacts. 

Many existing works on RF fingerprinting have relied on data collected using custom testbeds.  In the IoT space, several works have used LoRa and ZigBee tesbeds. In~\cite{shen_radio_2021}, several signal representations were considered using signals from 25 LoRa devices captured over  several days. To consider the impact of channel, indoor and outdoor captures were performed using 100 LoRa transmitters in~\cite{al-shawabka_deeplora_2021}. 
In~\cite{shen_towards_2021}, the problem of identifying unseen devices was approached using 60 LoRa devices. Using a capture 54 ZigBee radios, the  multi-sampling convolutional neural network was proposed for RF fingerprinting~\cite{yu_robust_2019}.  These previous works have only considered a single USRP receiver. 

The impact of changing receivers on transmitter fingerprinting was studied in \cite{merchant_toward_2019},  where signals from 25 ZigBee devices were captured using 10 receivers over a single capture. To the best of the authors' knowledge, these datasets are not publicly accessible. In~\cite{elmaghbub_lora_2021}, RF fingeprinting was evaluated under several experimental scenarios  using  25 LoRa devices and 2 receivers with the dataset publicly available.  Other works have used USRPs to send signals following different protocols: in~\cite{al-shawabka_exposing_2020}, 20 USRPs sending WiFi signals were used to study the impact of the channel. Four USRPs sending multiple  signals (5G New Radio, LTE, WiFi) were fingerprinted in~\cite{reus-muns_trust_2020}. Although, the datasets used in \cite{al-shawabka_exposing_2020,reus-muns_trust_2020} are publicly accessible, they only considered one receiver. Besides, USRPs have fundamentally different hardware than commercial radios and thus their transmitter fingerprints do not necessarily represent commonly used radios. All the previous datasets included at most 100 transmitters.  The largest RF fingerprinting dataset reported in the literature is the DARPA RFMLS dataset~\cite{darpa_radio_nodate}; it consists of signals from 5117 WiFi Tx captured using a spectrum analyzer and 5000 ADS-B devices~\cite{jian_deep_2020}. This dataset was used by many works~\cite{robinson_novel_2021,jian_deep_2020,al-shawabka_exposing_2020}; however, it is not publicly accessible. In our previous work~\cite{hanna_openset_2020}, we considered a WiFi dataset using  163  WiFi transmitters captured in the Orbit testbed~\cite{orbit_2005}, but it only considered one receiver.

 In this work, we present the WiSig dataset, including 10 million WiFi packets transmitted by 174 off-the-shelf WiFi transimtters and captured by  41 USRP receivers in four different captures performed over a month. To the best of the authors' knowledge, this dataset is the largest publicly available dataset and the first to include such a large number of both transmitters and receivers. Our focus was  on making the dataset widely accessible and easy to use. Instead of only making the 1.4 TB raw dataset (\emph{Raw WiSig}) available, we also provide a 76.9 GB processed dataset (\emph{Full WiSig}) containing directly usable WiFi preambles\footnote{https://cores.ee.ucla.edu/downloads/datasets/wisig/}. Furthermore, we provide conveniently pre-packaged subsets only  a few GB in size. Using these subsets, in this paper, we showcase example uses of WiSig and identify some open problems in RF fingerprinting. The scripts to generate the results in this paper are publicly available. We also provide the scripts to process the raw signals, create custom subsets, along with the commands to replicate the capture in a private testbed. Our main contributions can be summarized as follows
 \begin{itemize}
 	\item We provide a large-scale WiFi dataset captured by  41 USRP receivers from 174 WiFi transimtters over four different captures performed within a month.
 	\item We showcase how the dataset can be used to study the performance of RF fingerprinting as we change the number of transmitters, receivers, days, and signals.
 	\item We identify some open problems in RF fingerprinting that can be addressed using this dataset.
 \end{itemize}
The rest of the paper is organized as follows: in Section~\ref{sec:overview}, we provide an overview of the dataset creation. The data capture, signal extraction, and dataset assembly are described in Sections~\ref{sec:raw_sig},~\ref{sec:extraction}, and~\ref{sec:assembly} respectively. Use cases of the dataset along with open problems are provided in Section~\ref{sec:usecases}. Section~\ref{sec:conclusion} concludes the paper.

\begin{table}[t]
	\renewcommand{\arraystretch}{1.5}
	\caption{Summary of Related Datasets \label{tbl:related work}}
	\centering
	\begin{tabular}{|p{0.3cm}|p{2cm}|p{2cm}|p{1cm}|p{1cm}|}
		\hline
		Ref                    &  Transmitter & Receiver &  Many Days & Public \\	\hline
		\cite{shen_radio_2021} & 25 LoRa      & 1 USRP       & Yes & No\\ 	\hline 
		\cite{al-shawabka_deeplora_2021} & 100 LoRa      & 1 USRP       & Yes & No\\ 	\hline 
		\cite{shen_towards_2021} & 60 LoRa      & 1 USRP       & No & No\\ 	\hline 
		\cite{yu_robust_2019} &54 ZigBee      & 1 USRP       & No & No\\ 	\hline 
		\cite{elmaghbub_lora_2021}  & 25 LoRa      & 2 USRP    & Yes & Yes \\ 	\hline
		\cite{merchant_toward_2019} &25 ZigBee      &10 USRP, Spectrum analyzer       & No & No\\ 	\hline 
		\cite{al-shawabka_exposing_2020} &20 USRP (WiFi signal)      &    1 USRP    & Yes & Yes\\ 	\hline 
		\cite{reus-muns_trust_2020} &4 USRP (signals: 5G New Radio, LTE, WiFi)      & 10 USRP, Spectrum analyzer       & No & Yes\\ 	\hline 
		\cite{darpa_radio_nodate} &5117 WiFi , 5000 ADS-B      &  RF Spectrum analyzer       & Yes & No\\ 	\hline 
	\end{tabular} 
\end{table}
%

%
%
%
%
%
%
%
%

%
%
%
%
%
%
%
%

\section{Dataset Overview}
\label{sec:overview}
\begin{figure}[t!]
	\centerline{\includegraphics[scale=1]{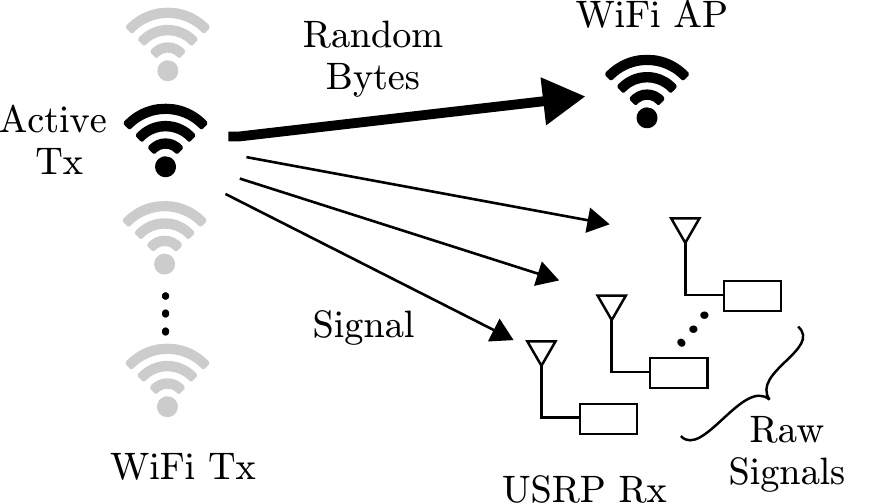}}
	\caption{The capture setup. The active Tx sends random bytes using UDP  to a WiFi access point. The USRP Rx capture the signals over the same band. }
	\label{fig:data_capture}
\end{figure}
	\begin{figure}[t!]
	\centerline{\includegraphics[scale=1]{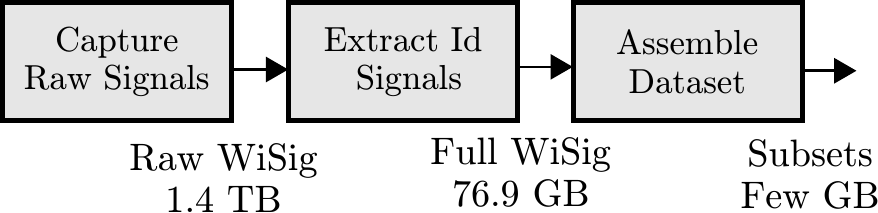}}
	\caption{ The process for creating the dataset. Raw signals are captured, then the identification signals are extracted, and finally the dataset is assembled. The size of the data decreases after each stage.}
	\label{fig:high_level_process}
\end{figure}
The dataset consists of USRP captures of signals sent from a WiFi node to a WiFi access point (AP); the WiFi nodes are the transmitters to fingerprint, the USRPs are the receivers, and the WiFi AP is needed to establish the WiFi link.  
The capture from one WiFi transmitter works as follows;  a transmitter sends data to the AP while all the USRP receivers continuously capture the raw IQ samples as illustrated in Fig.~\ref{fig:data_capture}.  In a single day, captures were performed from all contributing transmitters, one at a time, in a fully automated manner. The raw IQ samples for the entire duration of the transmission are stored, including idle time.  Four single-day captures are performed over a month to create the Raw WiSig dataset. The Raw WiSig dataset has a large size of 1.4 TB and needs to be processed before usage. 

We performed the processing to create the \emph{Identification (Id) signals} that can be easily used as input to a classifier. As processing, we detected the packets and  isolated them. After packet detection, 
two types of identification signals are created; the first one uses the first 256 IQ samples from each packet without further processing. In the second one, the packet is equalized, using the WiFi preambles, before extracting the first 256 IQ samples. All Id signals (non-equalized and equalized) constitute the Full WiSig dataset. The Full WiSig dataset size is 76.9 GB, which is still considerably large. Due to many sources of randomness in the capture, Full WiSig is not balanced; that is, not all transmitter-receiver pairs have the same number of signals. Additionally, a typical  user will probably only use a subset of the signals and might not need the entire Full WiSig.

For convenience, we assembled four subsets of the Full WiSig dataset to cater to the needs of different dataset users. These susbsets are called ManyTx, ManyRx, ManySig, SingleDay. ManyTx, ManyRx, ManySig provide a large number of Tx, Rx, or signals, respectively, covering all four days. SingleDay provides an appreciable number of Tx, Rx and signals on just one day. These susbsets are pre-packaged and only require a download of a few gigabytes. An overview of these steps is highlighted in Fig.~\ref{fig:high_level_process}. The scripts to process the dataset from one stage to the next one are provided, so that the user can modify each processing step  easily, if needed. We also provide scripts that enable the users to  repeat the capture setup using their own hardware. In the following sections, we describe each stage in detail.

\begin{figure}[t!]
	\centerline{\includegraphics[width=2.0in]{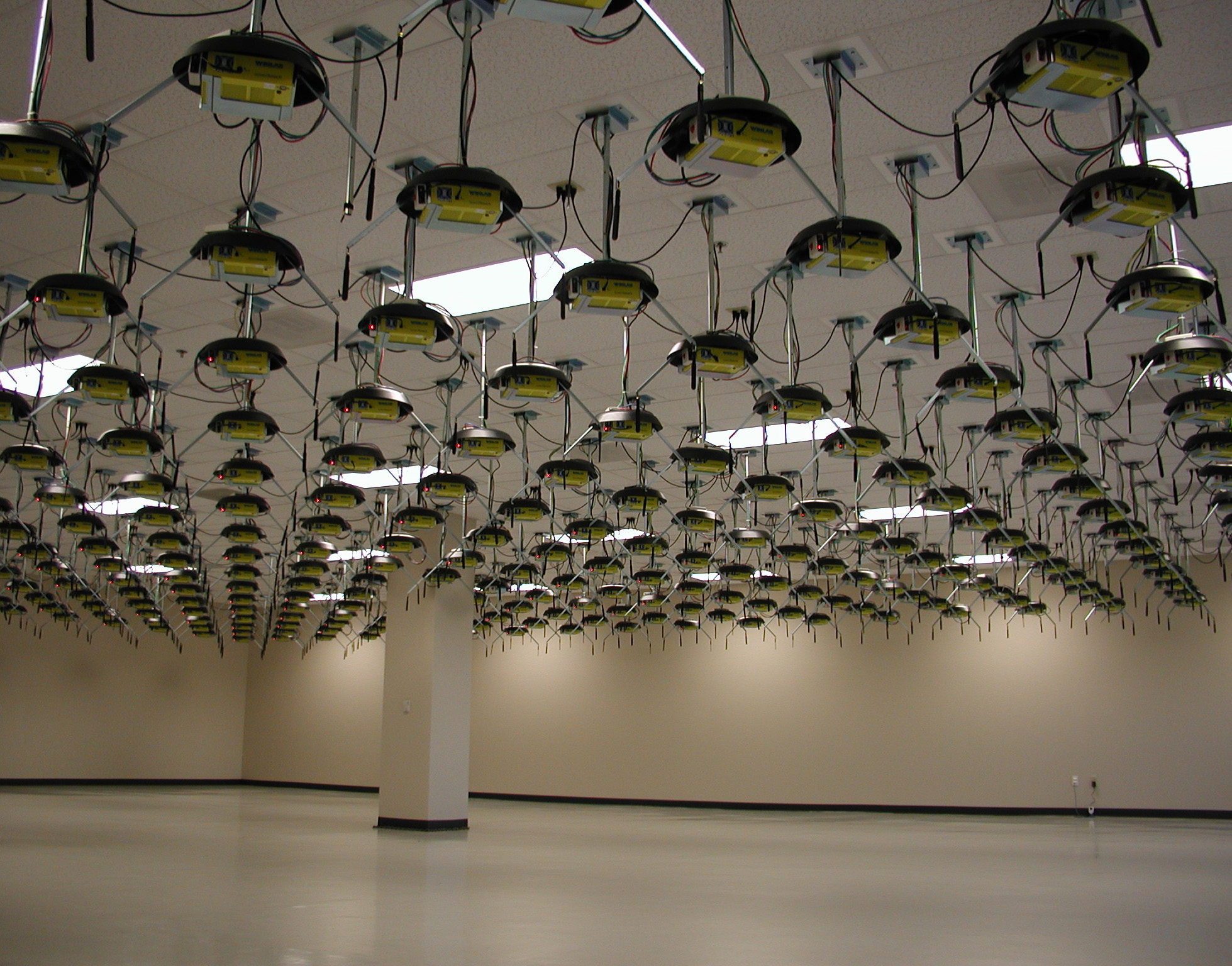}}
	\caption{Orbit nodes are arranged as a grid. Each node is a roof mounted PC with at least one WiFi module. Some nodes are equipped with USRPs.}
	\label{fig:orbit_grid}
\end{figure}
	\begin{figure}[t]
	\centering
	\subfloat[Transmitters \label{fig:tx_loc}]{\includegraphics{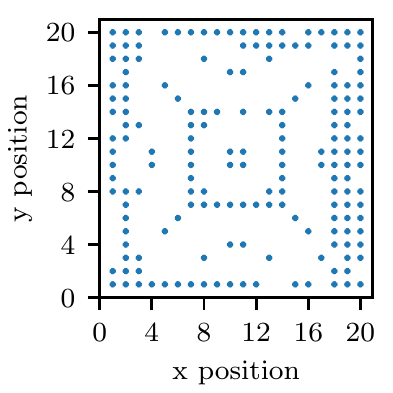}}  \hspace{5mm}
	\subfloat[Receivers \label{fig:rx_loc}]{\includegraphics{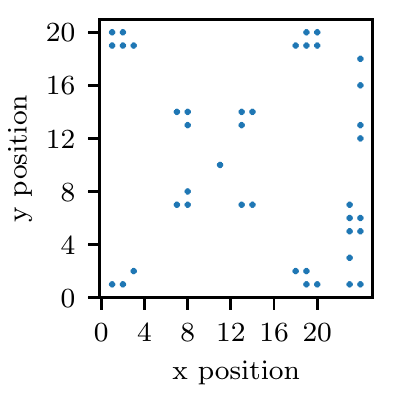}}  
	\caption{Positions of Tx and Rx in the Orbit grid. Rx with x position greater than 20 are part of the massive MIMO racks.}
	\label{fig:tx_rx_loc}
\end{figure}

\section{Capturing Raw Signals}
\label{sec:raw_sig}
The dataset was captured in the Orbit testbed~\cite{orbit_2005}. The Orbit testbed consists of a $20\times 20$ grid of nodes with a separation of 3 feet (about 1m). Each node is a roof-mounted computer equipped with at least one WiFi radio; some nodes are equipped with USRPs. An image of the Orbit grid is shown in Fig.~\ref{fig:orbit_grid}.  In addition to the grid, some USRPs are placed in the same room as part of two massive MIMO racks.  The locations of the WiFi transmitters and the USRP Rx are shown in Fig.~\ref{fig:tx_rx_loc}.  The WiFi 802.11 a/g modules used are Atheros 5212, 9220,  9280, and 9580; the USRPs are B210, X310, and N210. Note that the USRPs with x-position larger than 20 in Fig.~\ref{fig:rx_loc} are part of the massive MIMO racks. The captures were performed on four days: the 1st, 8th, 15th, and 23rd of March 2021. 

We start by describing the  WiFi transmitters (Tx), AP, and USRP receivers configurations. The WiFi 802.11 a/g Tx and AP were configured to operate over WiFi channel 13 having a center frequency of 2462MHz and a bandwidth of 20MHz. This channel was chosen because no non-participating WiFi APs were detected on this channel. However, the Orbit grid is not RF isolated and external interference is possible.  To avoid any data clues about the identity of the WiFi transmitter, all  transmitters were configured to have the same spoofed MAC address and the same IP address when connected. The WiFi payload  consisted of UDP packets carrying a stream of random bytes. The USRP receivers captured IQ samples at a rate of 25 Msps with a center frequency of 2462MHz for a duration of 0.512s. Due to the mostly line-of-sight channel, the signals are at high SNR (>10dB).  Note that the number of WiFi packets transmitted within the capture duration is determined by the WiFi MAC protocol and hence it varies per capture. Also note that   all USRP receivers operated independently without any time or frequency synchronization. Due to the delays in processing the commands, there is no guarantee that the captures are aligned in time; that is some receivers might start the capture earlier or later than others. 

The number of WiFi transmitters that participated in all the captures was  174 and the number of USRP receivers was 41. However, not all transmitters or receivers participated in all days of the data capture for several reasons; (1) On some days, some transmitters or receivers were not accessible in the Orbit testbed (2) A transmitter or receiver can run into hardware or software problems on one or multiple occasions.  Due to the large scale of the experiment, it was not feasible to stop the experiment and manually debug each transmitter or receiver.  As a consequence, there is no data for some transmitter-receiver pairs in the raw dataset on some days.  Note that if the transmitter did not operate, the USRPs would still capture a signal and this case needs to be handled in processing. 

\section{Extracting Identification Signals}
\label{sec:extraction}
	\begin{figure}[t!]
	\centerline{\includegraphics[scale=1]{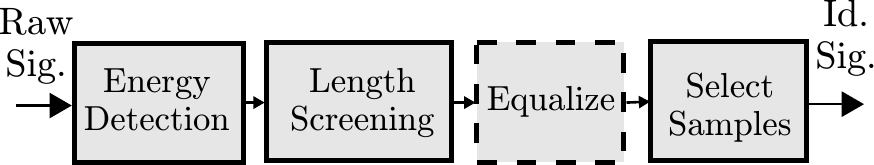}}
	\caption{The processing flow for extracting the identification signals from the raw signals. Two types of Id signals are created, one with and the other without equalization.}
	\label{fig:signal_extract}
\end{figure}
To obtain the Id Signals (both non-equalized and equalized), we start by detecting the packets and eliminating irrelevant signals. Then we perform equalization for one version of the Id signal, then we choose the samples to be used as illustrated in Fig.~\ref{fig:signal_extract}.

\subsection{Energy Detection and Screening}
In this stage, our objective is to identify the IQ samples corresponding to transmissions from the WiFi nodes and exclude other signals in the spectrum like the ACK response from the WiFi AP and non-WiFi signals that might exist.  We start by detecting signals and then excluding the ACKs.   The signal detection was performed by comparing the signal magnitude in a window consisting of $N_w$ IQ samples to a fixed threhold $L_w$. As for identifying the packets from the ACK, we relied on signal durations and the fact that each  WiFi packet is much longer than  the fixed duration ACK signal that follows it. We considered a valid packet as  a signal of length larger than $N_{pkt}$ followed immediately by a signal shorter than $N_{ack}$.   Both the packet detection and isolation were performed  without using the known WiFi preambles.  While this is not ideal from a detection perspective,  it is simple to implement and emulates the processing for protocol agnostic transmitter fingerprinting, even though we do not consider it in this work. We used  $N_w=100$, $L_w=0.005$, $N_{pkt}=1000$, and $N_{ack}=2000$ in our detection script. These values were obtained empirically by visually inspecting the captured IQ samples. After isolating the packets, for some transmitters only a few packets were detected, which might indicate that the transmitter did not turn on properly and these packets were erroneously detected. So, transmitters with  exceptionally few captured packets (specifically, less than 10) were eliminated.

\subsection{Equalization and Selecting Samples}
Using the isolated packets, two datasets were built; the first one contains part of the unprocessed preambles, the second includes equalization. For the first dataset, the first 256 IQ samples of each packet were included along with the transmitter, receiver, and day labels. In the second dataset, the  preambles were equalized. To equalize the packets, we used the WiFi packet structure, which starts with two preambles: the L-STF and L-LTF preambles~\cite{perahia_next_2013}. First, we resampled the packet to change its sampling rate from 25 Msps to 20 Msps because it is the nominal sampling rate for WiFi.  Then  autocorrelation is applied on the L-STF preamble to accurately detect the packet start. If the preamble is detected the frequency offset is estimated and corrected, else the packet is discarded.  The L-STF preamble misdetection can be due to the misidentification the start of the signal using energy detection  or caused by a mistakenly captured a non-WiFi signal. For the detected packets, the channel is then estimated using the L-LTF and the signal is equalized using MMSE. The signal processing for detection and channel estimation was applied using MATLAB R2019b WLAN toolbox with the default parameters. Afterwards, the frequency offset is reapplied to the signal to be used for fingerprinting. The signal is then resampled back to 25Msps and the first 256 samples were included in the dataset along with the transmitter and receiver labels. After equalization, the transmitters with few packets were eliminated again. The scripts to perform these steps are provided with the downloads.

\section{Dataset Assembly}
\label{sec:assembly}
	\begin{figure}[t!]
	\centerline{\includegraphics[scale=1]{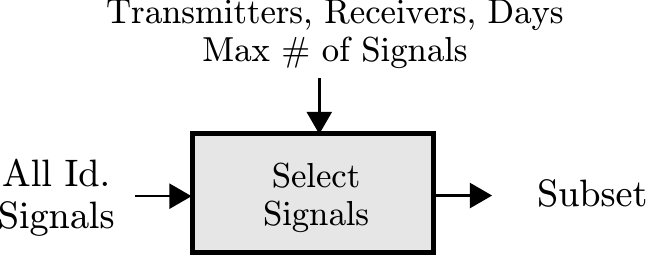}}
	\caption{The dataset assembly flow. A subset is created from all the identification signals by specifying the list of Tx, Rx, days, and the maximum number of signals to retain.}
	\label{fig:assemble_dataset}
\end{figure}
After capturing and processing the data, a WiSig user would typically choose a subset of the dataset that meets the users requirements. This subset is chosen by providing a list of Tx, Rx, days, and the maximum number of signals as illustrated in Fig.~\ref{fig:assemble_dataset}. We start by analyzing the number of signals and present a method to select a subset of the dataset. We also discuss the prepackaged  subsets provided for the convenience of WiSig users.

\subsection{Dataset Analysis}
\begin{figure*}[t!]
	\centerline{\includegraphics[scale=1]{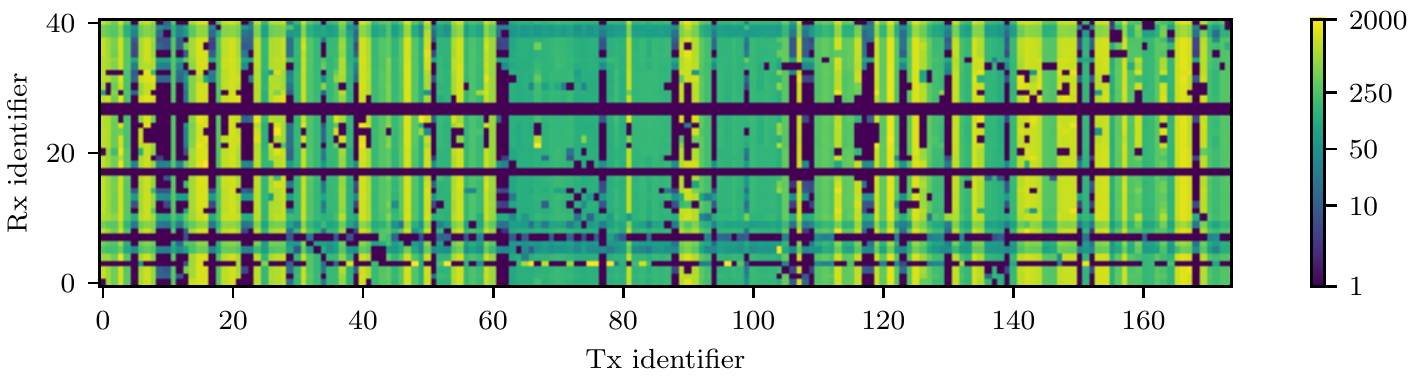}}
	\caption{The signal count for each Tx-Rx pair on the first day using a logarithmic colormap. Some Rx did not capture any signals for that day. Due to the fact that Rx are not synchronized and possibly due to detection errors, the signal count from different Rx for the same Tx can be significantly different.}
	\label{fig:tx_rx_grid_count}
\end{figure*}
\begin{figure}[t!]
	\centerline{\includegraphics[scale=1]{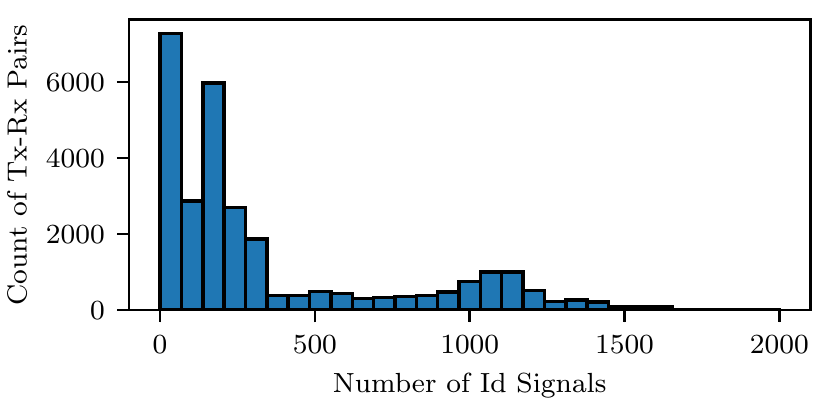}}
	\caption{Histogram showing the count of Tx-Rx pairs having a given number of signals. The histogram is calculated over all days for the non-equalized Full WiSig.}
	\label{fig:histogram}
\end{figure}
After processing the dataset, we analyze the number of collected Id signals $C(d,t,r)$ on  day $d$, from transmitter $t$, and receiver $r$. The total number of Id signals, that is  $\sum_{d,t,r} C(d,t,r)$, is equal to 9.97 million using 174Tx, 41 Rx, over 4 days. As discussed earlier, some transmitters and receivers failed on some days. Also, the number of WiFi packets transmitted varied due to the WiFi MAC rate control. Additionally, due to the lack of synchronization along with detection errors, the number of Id signals varied among Rx. 
All these factors make the dataset imbalanced as the number of Id signals $C$ depend on $d$, $t$, as well as $r$. To visualize $C$, in Fig.~\ref{fig:tx_rx_grid_count}, we plot $C(d,t,r)$ for the first day $d=0$. The x and y axis are identifiers for the particular Tx and Rx respectively, and the logarithmic colormap indicates the count of Id signals. To visualize all the counts, in Fig.~\ref{fig:histogram}, we plot the histogram of $C$ below 2000 for all  $d$, $t$, and $r$. From this figure, we see that the majority of Tx-Rx-day counts $C(d,t,r)$, is below 400. A smaller number of  Tx-Rx-day counts ($C$) exceeded 1000, and a few even exceeded 2000 but were not included in the histogram for clarity. Note that because the L-LTF preamble detection for some signals, the number of collected signals for the equalized signals $C_{\text{eq}}(d,t,r)$ might be slightly smaller than the non-equalized dataset, that is $C_{\text{eq}}(d,t,r)\leq C(d,t,r)$.

 The large imbalance of $C(d,t,r)$ is typically not desirable as it can confuse training and give misleading results. Depending on the application, the required number of needed Tx, Rx, or signals can be smaller than the ones available in the dataset. To reduce the dataset imbalance,  the user would typically start by choosing a subset of the dataset.

\newcommand{\mTsub}{\mathcal{T}}
\newcommand{\mRsub}{\mathcal{R}}
\newcommand{\mDsub}{\mathcal{D}}
\newcommand{\mRn}{M}
\newcommand{\mTn}{N}
\newcommand{\mSign}{K}

\begin{figure}[t]
	\centering
	\subfloat[ For $p=1.0$, $M$ is obtained using the greedy heuristic.  \label{fig:max_rx_p1}]{\includegraphics{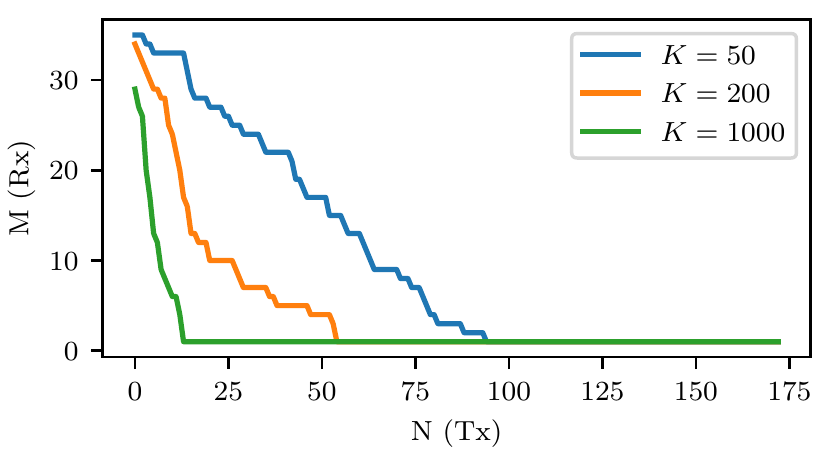}}  \hspace{5mm}
	\subfloat[For $p=0.9$, $M$ is obtained using the MILP solver. \label{fig:max_rx_p2}]{\includegraphics{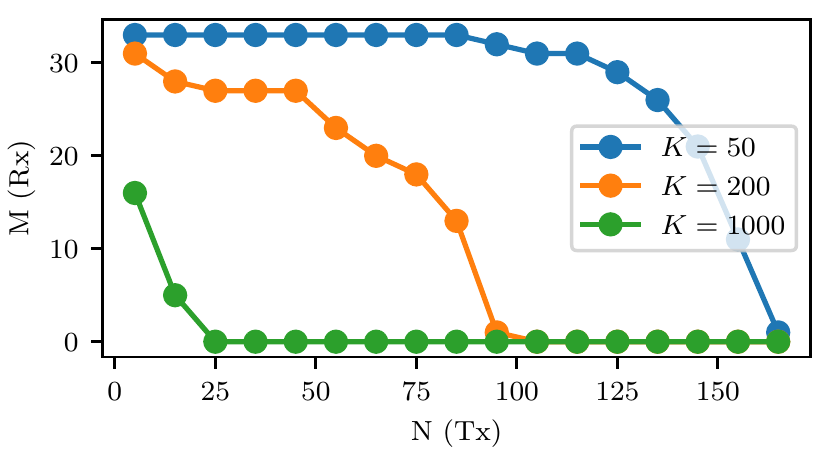}}  
	\caption{The number of Rx ($M$) for a given number of Tx ($N$) in WiSig, such that at least a fraction $p$ of Tx per Rx has at least $K$ signals. Using $p=0.9$ utilizes a larger number of Rx than $p=1$.}
	\label{fig:max_rx}
\end{figure}

\subsection{Creating Subsets}
To choose a subset, we need to specify a set of $\mTn$ transmitters $\mTsub$ and a set of  $\mRn$ receivers $\mRsub$. The choice of a subset determines the minimum number of signals per Tx-Rx pairs $\mSign$.    The variables $\mTn$, $\mRn$, and $\mSign$ are dependent; that is  we can specify only two variables  and the third one is upper bounded by the dataset.  To make it easier for WiSig users to choose the transmitter and receiver subsets, we provide a utility which maximizes the number of Rx such that at least a fraction $p$ of  $\mTn$ Tx have at least $\mSign$ signals. This problem can be formulated as follows for a given day $d$
\begin{align}
& \underset{\mTsub,\mRsub}{\text{maximize}} 
& &  |\mRsub| \label{eq:tx_rx_obj} \\
& \text{subject to}
& &  \sum_{t \in \mTsub} I\left[ C\left(d,t,r\right) \geq \mSign  \right]  \geq p |\mTsub| \qquad \forall \   r\in \mRsub \label{eq:tx_rx_const} 
\end{align}
where  $I[\cdot]$ is the indicator function which takes a value of one if the constraint is satisfied and $|\cdot|$ denotes the size of the set, thus $|\mTsub|=\mTn$ and $|\mRsub|=\mRn$.

This problem can be formulated as a mixed integer linear programming (MILP). While optimal, the MILP solution can be time consuming. So we developed a simple greedy algorithm  which gave satisfactory results when $p=1$. The MILP formulation is discussed in the appendix. The greedy heuristic works as follows:  we start by choosing the Tx having the most Rx  that have at least $\mSign$ signals. For these transmitters, we choose the receivers satisfying the constraint~(\ref{eq:tx_rx_const}). 

The values of $\mRn$ obtained using the greedy algorithm is shown in Fig.~\ref{fig:max_rx_p1} when considering the minimum over all days, that is $C'(t,r)=\min_d C(d,t,r)$ for $p=1$. Fig.~\ref{fig:max_rx_p1} shows that, as expected, number of increasing either the required transmitters $\mTn$ or $\mSign$ yields a smaller number of Rx $\mRn$. The problem with using $p=1$ is that even a single Tx-Rx pair not having $K$ signals would eliminate that Rx entirely. To relax this constraint, we can use $p=0.9$, for example; that is, only 90\% of Tx should satisfy the signal requirement per Rx. For $p=0.9$, we used the MILP formulation because the heuristic did not yield acceptable results. The MILP results are shown in Fig.~\ref{fig:max_rx_p2}; note that a larger portion of the dataset can now be utilized. 

\subsection{Prepackaged Compact Subsets}
\begin{table}[t]
	\renewcommand{\arraystretch}{1.5}
	\caption{Description of Compact Datasets \label{tbl:compact_subsets}}
	\centering
	\begin{tabular}{|c|c|c|c|c|c|}
		\hline
		Name & $\mTn$ (Tx) & $\mRn$ (Rx) & $\mSign$ (Sig) & $p$ & Days \\	\hline
		ManySig & 6 & 12 & \textbf{1000} & 1.0 & 4 \\ 	\hline 
		ManyTx & \textbf{150} & 18 & 50 & 0.9 &  4\\	\hline 
		ManyRx & 10 & \textbf{32} & 200 & 0.9 & 4\\	\hline 
		SingleDay & 28 & 10 & 800 & 1.0 & \textbf{1}\\	\hline 
	\end{tabular} 
\end{table}
To eliminate the need to download the full dataset just to use a subset, we provide conveniently pre-packaged compact subsets. We provide 4 subsets as shown in Table~\ref{tbl:compact_subsets}. These subsets were designed to cater to different possible use cases. The first subset ManySig was designed to be balanced and provide 1000 signals  for all Tx-Rx pairs over all days. ManyTx focuses on increasing the number of transmitters and provides 150 Tx, while tolerating a slight imbalance ($p=0.9$). ManyRx provides signals captured by a relatively large number of 32 Rx with a slight imbalance. All the previous subsets include data from all 4 days. For users not interested in the impact of days (channel variation), we created the SingleDay subset, which provides a relatively large number of signals and Tx but only for one day. The required download sizes for different versions of the dataset are shown in Table.~\ref{tbl:download_size}. Note that whereas the raw signal dataset would have required a tremendous 1.4 TB download, the size gets reduced to 76.9 GB after processing, while the pre-packaged subsets are a few GB each, which is more convenient. 

\begin{table}[t]
	\renewcommand{\arraystretch}{1.5}
	\caption{WiSig Download Size \label{tbl:download_size}}
	\centering
	\begin{tabular}{|c|c|c|}
		\hline
		Name & Signal & Download Size \\	\hline
		Raw WiSig & Raw & 1.4 TB \\	\hline
		Full WiSig & Processed & 76.9 GB  \\	\hline
		WiSig ManySig & Processed & 1.4 GB \\ 	\hline 
		WiSig ManyTx & Processed &  2.5 GB\\	\hline 
		WiSig ManyRx & Processed & 1.2 GB\\	\hline 
		WiSig SingleDay & Processed & 1.0 GB \\	\hline 
	\end{tabular} 
\end{table}

\section{Use Cases \& Open Problems }
\label{sec:usecases}
The purpose of this section is to demonstrate a few of the possible use cases of WiSig and highlight some of the open problems in transmitter identification. The evaluation setup considered is simple and the results provided are meant to identify problems and not to solve them.
We only considered the problem of closed-set classification among a fixed predetermined set of transmitters, even though it is not the only possible formulation. 
 
 Unless otherwise stated, the experimental setup was the following: we used the same neural network consisting of 5 convolutional layers--- each followed by max pooling---with three dense layers at the end. The convolutional layers used 8, 16, 32, and 16 filters of sizes (3,2), (3,2), (3,2),(3,1), and (3,1) respectively. The dense layers used 100, 80, and $\mTn$ units. All layers used ReLU activation except the output layer, which used softmax. The network was trained for 100 epochs with early stopping if the validation loss did not decrease for 5 epochs, and the best weights were retained. The loss function used is categorical crossentropy. Each dataset was divided as 80\% for training 10\% for validation and 10\% for testing. All signals were normalized to have unit  average power prior to using them. 
 
 We now move on to studying the generalization across receivers, days, and the impact of the number transmitters, and number of signals on classification. We also show that WiSig can be used for localization.

\subsection{Generalization across Receivers}
\begin{figure}[t!]
	\centerline{\includegraphics[scale=1]{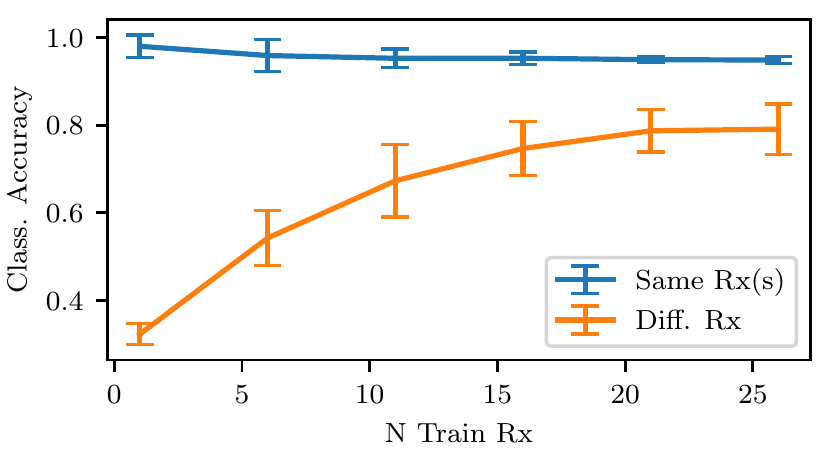}}
	\caption{The classification accuracy  as a function of the number of  receivers used in training. A  neural network is trained for each number  of Rx and is evaluated using data from the same training Rx and then using 5 different Rx. The equalized data is used from the ManyRx subset having 10 Tx.}
	\label{fig:nrx}
\end{figure}
Many wireless deployments involve multiple access points, and thus require transmitter authorization to be deployed on many receivers.  Therefore, it is important to investigate the impact of the receiver fingerprint on  classification accuracy. To do that, we use the ManyRx dataset, having 32 receivers. To reduce the impact of channel  variation associated with changing the receivers, we used the equalized dataset and just  a single day. We randomly chose a subset of these receivers and  trained a Tx classifier using their data. The number of receivers in this subset was varied and for each size we trained a classifier. Then, we evaluated each trained classifier using two test sets: the first one uses data from the same receivers as those used in training ("Same Rx") and the other one uses 5 entirely different receivers not used during training ("Diff. Rx"). The purpose of this division is to investigate whether a trained classifier can be used on a different receiver than those used to collect the training data. To increase the confidence in the results, we used 5 random realizations of the receiver sets, and show the results as mean and standard deviation in Fig.~\ref{fig:nrx}. Looking at the "Same Rx" curve, increasing the number of Rx slightly degrades the accuracy, which is expected, as increasingly more Rx fingerprints need to be learned. However, the average accuracy remains above 95\%. When we use  different receivers for evaluation, the performance is impacted significantly: for example, using only one Rx for training results in a drop in accuracy from 99\% to less than 33\%. This huge drop shows that a transmitter classifier trained on one Rx is not expected to work well on another Rx due to the impact of the receiver fingerprint. As we increase the number of Rx in training, the testing accuracy on different Rx improves, showing that including captured signals from  more Rx during training can improve generalization to new Rx. However, the improvement starts to saturate at about 20 Rx.

While using many Rx in the training capture is simple, it might not always be feasible; a large number of Rx might not be available during training. Ideally, we want to capture the Tx fingerprint using only one Rx and apply it to any Rx. Since the receiver fingerprint is also due to manufacturing variability and is hard to model,  data driven solutions are needed. One solution is to develop Rx fingerprint augmentation techniques. Once trained, such an approach  would use signals from one Rx to synthesize signals from other virtual receivers similar to the approach used in~\cite{karunaratne_open_2021}. Another approach is to develop a technique that can extract receiver independent features. Either one of these approaches could be developed and evaluated using  WiSig. 

\subsection{Generalization across Days}
\begin{figure}[t!]
	\centerline{\includegraphics[scale=1]{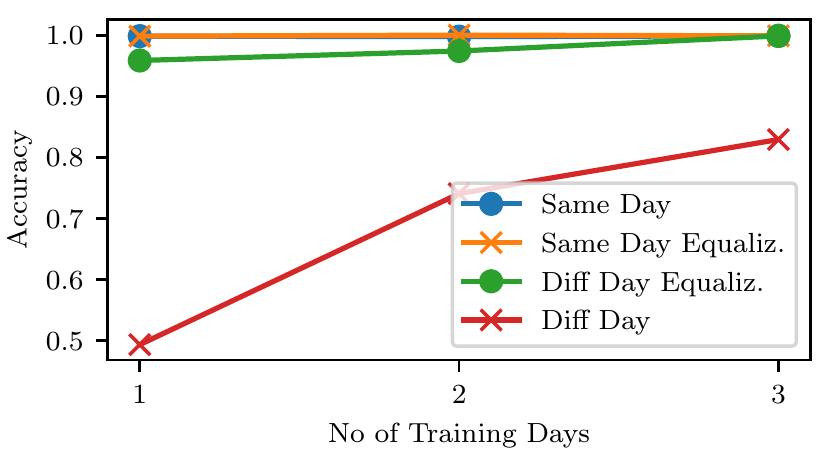}}
	\caption{The classification accuracy  as a function of the number of days used in training. Two  neural networks are trained for each number  of days: one for the non-equalized and one for the equalized data. Each networks is  evaluated using data from the same training days and a different day. The ManySig subset, having 6 Tx, is used.}
	\label{fig:ndays}
\end{figure}
Any real-life transmitter authorization deployment is expected to work over a prolonged period of time. As such, we use WiSig to evaluate the performance of  a trained transmitter classifier over time. To do that, we use the data from a single receiver from the ManySig dataset; the last day is reserved for testing and the first three days are used for training. We trained several networks using the data from either one, two or all three days and using both the non-equalized and equalized datasets. Then we evaluated them using data from the same days used in training as well as from the different day. The results are shown in Fig.~\ref{fig:ndays}. When the evaluation is performed on the same day (both training and testing data collected in the same 0.5s capture), accuracies exceeding 99\% are obtained.  However, when tested on the different day data (captured a few weeks later) the accuracy drops as previously demonstrated in~\cite{al-shawabka_exposing_2020}. As expected, including data from more days during training improves generalization. Equalizing the dataset reduces the channel variability, which decreases the performance degradation on a different day. However, there is still a small drop in accuracy for different days even with equalization, which might be because some aspects of the Tx fingerprint or Rx fingerprint may vary slightly over time~\cite{shen_radio_2021}. 

\begin{figure}[t!]
	\centerline{\includegraphics[scale=1]{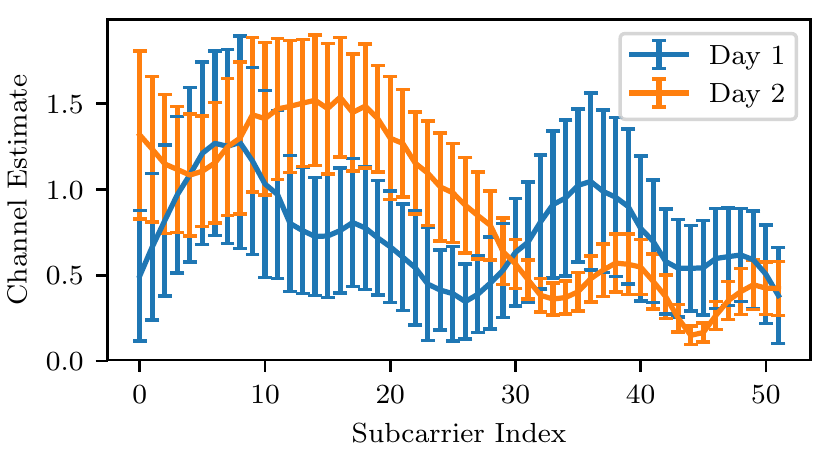}}
	\caption{ The channel estimates of the 52 occupied subcarriers over 2 days. The curve represents the mean over all packets per day and the errors bars represent the standard deviation. The packets are power-normalized before channel estimation.}
	\label{fig:channel}
\end{figure}
To better understand  the difference between signals captured on different days, we compare the channel estimates from two different days. To that end, we consider a single transmitter (node 1-1) and a single receiver (node 20-20). For all the packets from this Tx-Rx pair over two days (March 1st and 8th), after per-packet power normalization, we calculated the magnitude of the channel for all the 52 occupied WiFi subcarriers. In Fig.~\ref{fig:channel}, we plot the mean and standard deviation of the channel calculated over all packets for each day; we can see clearly  that the channel differs significantly between the two captures, even though we used the same Tx and Rx. This difference can be explained by changes in the propagation environment (for example people moving objects), which occurred during the week separating the two captures.

Of course, capturing data  over a prolonged period of time for robust transmitter identification is not desirable. Ideally, we only want to isolate the Tx fingerprint using just one capture of a short duration. To achieve that goal, we can use channel augmentation techniques, whether handcrafted or data-driven similar to those proposed in~\cite{soltani_more_2020}.  Another approach is to use neural network architectures robust to channel variations similar to~\cite{brown_charrnets_2021}.  Further investigation is needed to understand how both days and receivers impact performance and whether using additional receivers can compensate for channel variation along days. WiSig  provides the data needed for these studies.

\subsection{Impact of Number of Training Signals}
\begin{figure}[t!]
	\centerline{\includegraphics[scale=1]{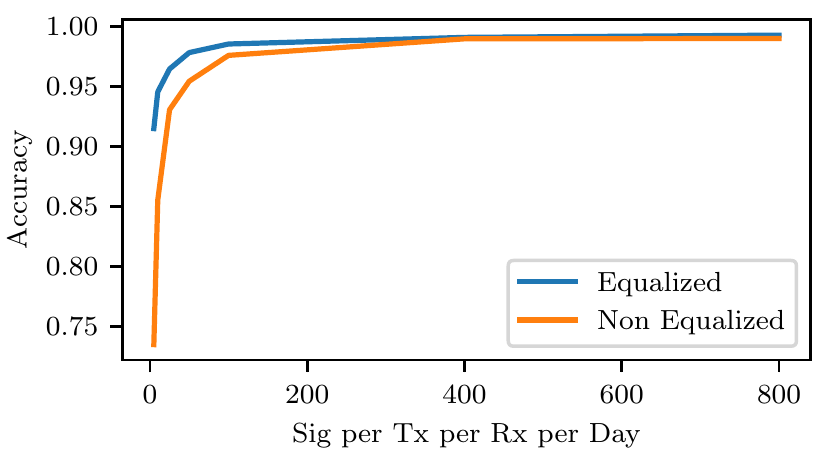}}
	\caption{The classification accuracy  as a function of the number of signals per Tx per Rx and per day. A  neural network is trained for each point. Data from all days and receivers in the ManySig subset is used for both training and evaluation. }
	\label{fig:nsig}
\end{figure}
Another aspect to consider is the number of signals needed to train a transmitter identification system and whether we should equalize signals or not. The WiSig dataset can be used to address these questions. Using the ManySig dataset, we train multiple classifiers each using a different number of signals per Tx per Rx per day.  We consider all 4 days and all 12 Rx in ManySig and we repeat the experiment with the non-equalized and the equalized datasets. Both the training and test signals are from the same Tx, Rx, days, and state of equalization. The results are   shown in Fig.~\ref{fig:nsig}. For a small number of signals, equalizing the dataset significantly improves the performance because it reduces the randomness from the channel and makes the task easier. As we increase the number of signals, accuracy of both approaches improves and the gap between them diminishes. With enough signals from all days and Rx, classifiers can counteract the randomness from the channels. 

Clearly, having more signal improves classification performance.  However, that might not always be feasible and neural network architectures that are data-efficient are needed. The minimum number of signals needed for different number of receivers, and days is also still an open problem~\cite{oyedare_estimating_2019}. Currently, the equalization is implemented using signal processing using the known WiFi preambles.  However, a neural network that can blindly equalize the signal for fingerprinting is desirable. Additionally, having a large number of signals can enable researchers to better evaluate the robustness of transmitter fingerprinting against sophisticated adversarial attackers~\cite{karunaratne_penetrating_2021}. The WiSig dataset can be used to address these open problems.

\subsection{Impact of Number of Transmitters}
\begin{figure}[t!]
	\centerline{\includegraphics[scale=1]{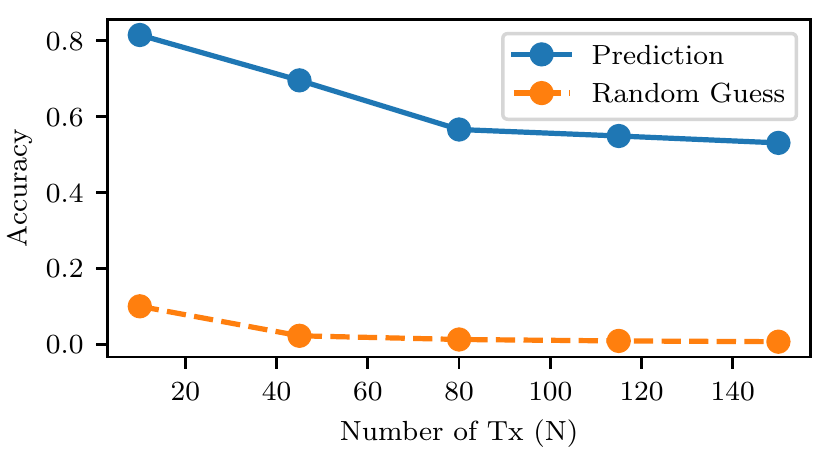}}
	\caption{The classification accuracy  as a function of the number of transmitters $\mTn$. A  neural network is trained for each point. Non-equalized data from all days and receivers in ManyTx is used for both training and evaluation. The random guess corresponds to $1/\mTn$.}
	\label{fig:ntx}
\end{figure}
Depending on the application, a transmitter authorization system might need to authenticate hundreds of users. WiSig enables the evaluation of transmitter authorization at this large scale of users with the ManyTx dataset containing signals from 150 transmitters. To showcase this dataset, we evaluate the impact of the number of transmitters $\mTn$ on classification accuracy. Considering all days and receivers, for each $\mTn$, we train a classifier and evaluate it on the same non-equalized data. The results are shown in Fig.~\ref{fig:ntx}. When using 4 days, 19 Rx, and only 50 signal per Tx per Rx per day, the classification accuracy with just 10 Tx is around 80\%. The accuracy is not very high because of the limited number of signals and the challenging structure including many Rx and days. As we increase the number of Tx, the problem becomes more challenging and the accuracy drops to about 53\% with 150 Tx. This highlights the need for neural network architectures that can perform well for a large number of transmitters.

Having many transmitters can enable using more practical formulations of the problem. For instance, instead of classifying among a closed set of known transmitters, open set recognition enables rejecting unauthorized transmitters unseen in training~\cite{hanna_openset_2020}. The challenge in developing these systems is that different Tx are needed for training and testing, as unauthorized transmitters cannot be exposed during training. Thus a large number of Tx is needed. Furthermore, by encoding the fingerprints to vectors and  inserting them into a database, fingerprinting can also become a lookup operation similar to what was proposed in~\cite{karunaratne_real-time_2021,shen_towards_2021}. Thus it is clear that the many transmitters provided by WiSig enables the development of novel architectures and exploring different ways to pose the authorization problem.

\subsection{Localization using Multiple Receivers}
\begin{figure}[t!]
	\centering
	\subfloat[Transmitter is node 2-3. \label{fig:loc_0}]{\includegraphics[scale=1]{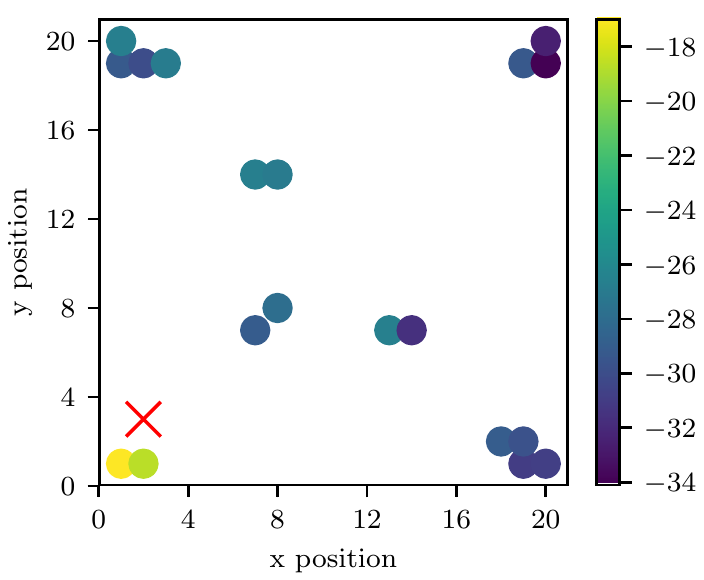}}\\
	\subfloat[Transmitter is node 13-7. No data from Rx  14-7. \label{fig:loc_1}]{\includegraphics[scale=1]{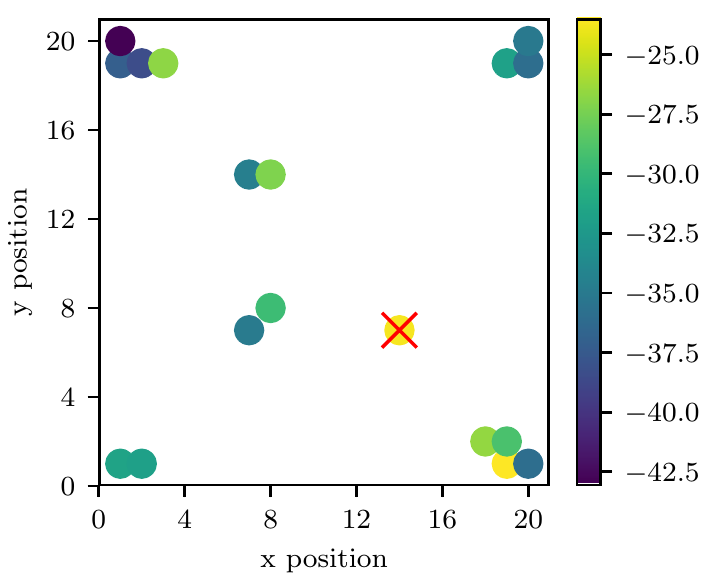}}
	\label{fig:loc}
	\caption{Average baseband received power from two different transmitters as seen by all receivers. Transmitters are shown as red crosses and receivers as circles. The Rx color indicates the average baseband received power in dBm. The ManyTx dataset is used.}
\end{figure}
Although the WiSig dataset was developed for RF fingerprinting,  the grid structure of transmitters and receivers enables using WiSig for localization.  To demonstrate that, for each transmitter, we calculated the average baseband power for all its packets as received by different Rx on one day. Using the ManyTx subset,  the results are shown in Fig.~\ref{fig:loc_0} and~\ref{fig:loc_1} for two example transmitters shown as a red cross. The 18 Rx are plotted as circles with the power represented by the heatmap in dBm. As expected, receivers closer to the transmitter have a higher power than those further away. This power received by all Rx creates a location fingerprint, which we removed in the previous results by normalization. To demonstrate that the Rx power can be used for localization, we trained a neural network using the location fingerprint consisting of 18 power measurements. For the missing Tx-Rx pairs, to avoid providing clues to the neural network, a random number within the maximum and minimum receiver power is used instead. The network consisted of three dense layers, consisting of 100, 80, and 2 units respectively. It was trained to predict the x-y coordinate of the transmitter using the mean squared error loss. The trained network provided an L1 error of 1.5 m on the test set. Since the grid separation is 1 m, each transmitter can be confused with its 2 nearest neighbors and only 8\% of the 150 Tx were accurately identified on the grid using their locations. 
While WiSig might not be the best dataset for localization research, it can enable combining localization and Tx fingerprinting. Localization using power can be used to estimate the location of the transmitters and fingerprinting can identify them at much higher accuracy than just relying on localization.





\section{Conclusion}
\label{sec:conclusion}
In this paper, we presented WiSig as a large scale dataset for transmitter fingerprinting. A preliminary evaluation using WiSig demonstrated the detrimental impact of changing training days, receivers, and limited training signals on performance. Although including more days, receivers, training signals, or equalizing leads to improved generalization, it might not always be feasible. Novel approaches are needed to develop more robust fingerprinting systems with less strict training data requirements. Further analysis is needed to fully understand the combined effects of  training days, and number of receivers, transmitters, along with equalization  on performance and to specify the required data collection requirements for good performance.   By highlighting  some of the open problems and making the data publicly available in an easily accessible manner, we aim to empower future research in transmitter fingerprinting.
\appendix
\section{Appendix}

A more general version of the problem in \eqref{eq:tx_rx_obj}-\eqref{eq:tx_rx_const} was used for the MILP formulation found in the accompanying helper code, and is presented below. $\mRsub$ and $\mTsub$ respectively represent the subsets of chosen transmitters and receivers, while $\mDsub$ represents the set of all days. $C_{\text{eq}}$ is the counterpart to $C$ that contains the equalized signals.  $\mRsub$ and $\mTsub$ were chosen such that the specified requirements were met across all days and across both the non-equalized and equalized data.
\begin{align*}
& \underset{\mTsub,\mRsub,k}{\text{maximize}}
& & w_{\infty} \mRn + k \\
& \text{subject to}
& &  \sum_{t \in \mTsub} I\left[ C\left(d,t,r\right) \geq \mSign  \right]  \geq p \mTn \qquad \forall r \in \mRsub, d \in \mDsub  \\
&&& \sum_{t \in \mTsub} I\left[ C_{\text{eq}}\left(d,t,r\right) \geq \mSign  \right]  \geq p \mTn \qquad \forall r \in \mRsub, d \in \mDsub\\
&&& C\left(d,t,r\right) \ge k \qquad \forall r \in \mRsub, d \in \mDsub, t \in \mTsub  \\
&&& C_{\text{eq}}\left(d,t,r\right) \ge k  \qquad \forall r \in \mRsub, d \in \mDsub, t \in \mTsub  \\
&&& k \ge K_{\text{low}} \\
&&&|\mTsub|=\mTn
\end{align*}
For this implementation, the following requirements were imposed:
\begin{itemize}
    \item The user specifies $\mTn = |\mTsub|$, $K$ and $p$ as in \eqref{eq:tx_rx_obj}-\eqref{eq:tx_rx_const}. In addition, the user is also allowed to specify $K_{\text{low}}$: for any $r \in \mRsub, d \in \mDsub, t \in \mTsub$ where $C(d,t,r) < K$, $K_{\text{low}}$ imposes a lower bound by ensuring that $C(d,t,r) \ge K_{\text{low}}$. This essentially allows the user to specify a tolerance level for $K$. Note that in \eqref{eq:tx_rx_obj}-\eqref{eq:tx_rx_const},   $K_{\text{low}} = 0$.
    \item For $r \in \mRsub, t \in \mTsub$, let $k = \min_{t, r} C_{\text{min}}(t, r)$ where $C_{\text{min}}(t, r) = \min_{d} \min\left\{ C(d, t, r), C_{\text{eq}}(d, t, r) \right\}$. The objective is to find $\mRsub, \mTsub$ that first maximizes $\mRn = |\mRsub|$, and then maximizes $k$ such that $|\mRsub|$ is not reduced. This hierarchical optimization can be achieved by giving an extremely large weight $w_{\infty}$ to $M$.
\end{itemize}

The MILP formulation of the above problem is given below: 
\begin{align}
& \underset{
\substack{\mathbf{T},\mathbf{R},\mathbf{Y}\\ \mathbf{Z},\mathbf{Q} \\ 
\mathbf{Q}_{\text{eq}}, k}}{\text{max}}
& & w_{\infty} \cdot \mathbf{1}^{\top}\mathbf{R}  + k \\
& \text{s.t.}
& &  Y(t, r) \le T(t) \qquad \forall r \in \mathbf{R}, t \in \mathbf{T} \label{eq:y_1} \\
&&& Y(t, r) \le R(r) \qquad \forall r \in \mathbf{R}, t \in \mathbf{T}  \label{eq:y_2} \\
&&& Y(t, r) \ge T(t) + R(r) - 1 \qquad \forall r \in \mathbf{R}, t \in \mathbf{T}  \label{eq:y_3} \\
&&& Z(t, r) \le Y(t, r) \cdot U \qquad \forall r \in \mathbf{R}, t \in \mathbf{T}  \label{eq:z_1} \\
&&& Z(t, r) \ge 0 \qquad \forall r \in \mathbf{R}, t \in \mathbf{T}  \label{eq:z_2} \\
&&& Z(t, r) \le k \qquad \forall r \in \mathbf{R}, t \in \mathbf{T}  \label{eq:z_3} \\
&&& Z(t, r) \ge k - [1 - Y(t, r)] \cdot U \qquad \forall r \in \mathbf{R}, t \in \mathbf{T}  \label{eq:z_4} \\
&&& C_{\text{min}}\left(t,r\right) \cdot Y(t, r) \ge Z(t, r)  \qquad \forall r \in \mathbf{R}, t \in \mathbf{T}  \\
&&& Q(d, t, r) \le Y(t, r)  \qquad \forall r \in \mathbf{R}, d \in \mathbf{D}, t \in \mathbf{T}  \\
&&& Q(d, t, r) \le  \bar{C}\left(d,t,r\right) \qquad \forall r \in \mathbf{R}, d \in \mathbf{D}, t \in \mathbf{T}  \\
&&& Q(d, t, r) \ge Y(t, r) + \bar{C}\left(d,t,r\right) -1 \qquad \nonumber \\
    &&& \qquad  \qquad  \qquad  \qquad  \qquad \forall r \in \mathbf{R}, d \in \mathbf{D}, t \in \mathbf{T}  \\
&&& Q_{\text{eq}}(d, t, r) \le Y(t, r)  \qquad \forall r \in \mathbf{R}, d \in \mathbf{D}, t \in \mathbf{T}  \\
&&& Q_{\text{eq}}(d, t, r) \le  \bar{C}_{\text{eq}}\left(d,t,r\right) \qquad \forall r \in \mathbf{R}, d \in \mathbf{D}, t \in \mathbf{T}  \\
&&& Q_{\text{eq}}(d, t, r) \ge Y(t, r) + \bar{C}_{\text{eq}}\left(d,t,r\right) -1 \nonumber\\
	&&&	 \qquad  \qquad  \qquad  \qquad  \qquad   \forall r \in \mathbf{R}, d \in \mathbf{D}, t \in \mathbf{T}  \\
&&& \mathbf{1}^{\top} q(d, r) \ge p\cdot N \cdot R(r) \qquad \forall r \in \mathbf{R}, d \in \mathbf{D}  \\
&&& \mathbf{1}^{\top} q_{\text{eq}}(d, r) \ge p\cdot N \cdot R(r) \qquad \forall r \in \mathbf{R}, d \in \mathbf{D}  \\
&&& k \ge K_{\text{low}} \\
&&& \mathbf{1}^{\top}\mathbf{T} =\mTn
\end{align}

\begin{itemize}
    \item Let $\mathbb{B} = \{0, 1\}$. $\mathbf{T} \in \mathbb{B}^{N_T \times 1}$ and $\mathbf{R} \in \mathbb{B}^{N_R \times 1}$ are binary variables that denote which transmitters and receivers are selected, respectively, where The total number of transmitters is given by $ N_T=174$  and receivers $N_R=41$.
    \item \eqref{eq:y_1}-\eqref{eq:y_3} accomplishes $Y(t, r) = T(t)~\&~R(r)$ where $\mathbf{Y} \in \mathbb{B}^{174 \times 41}$ is a binary variable.
    \item \eqref{eq:z_1}-\eqref{eq:z_4} accomplishes $Z(t, r) = \begin{cases}
    0 & ;Y(t, r) = 0 \\
    k & ;Y(t, r) = 1 \\
    \end{cases}$ where $\mathbf{Z} \in \mathbb{Z}^{N_T \times N_R}$ is an integer variable. Note that $U$ is any upper-bound to $C(d, t, r)$ and $C_{\text{eq}}(d, t, r)$.
    \item $\bar{\mathbf{C}} \in \mathbb{B}^{N_D \times N_T \times N_R}$, $\bar{\mathbf{C}}_{\text{eq}} \in \mathbb{B}^{N_D \times N_T \times N_R}$ are binary constants and $N_D=4$ is the number of days. $\bar{C}\left(d,t,r\right) = \begin{cases}
    0 & ;C(d, t, r) < K \\
    1 & ;C(d, t, r) \ge K \\
    \end{cases}$. $\bar{\mathbf{C}}_{\text{eq}}$ is defined similarly. 
    \item $\mathbf{Q} \in \mathbb{B}^{N_D \times N_T \times N_R}$, $\mathbf{Q}_{\text{eq}} \in \mathbb{B}^{N_D \times N_T \times N_R}$ are binary variables. $Q(d, t, r) = \begin{cases}
    1 & ; T(t)~\&~R(r)~\&~I[C(d, t, r) \ge K]  \\
    0 & ; \text{otherwise}\\
    \end{cases}$. $\mathbf{Q}_{\text{eq}}$ is defined similarly.
    
    \item $q(d, r) \in \mathbb{B}^{N_T \times 1}$ is a slice of $\mathbf{Q}$ where $q(d, r) = \{Q(d, t, r)~|~t \in \mathbf{T}\}$. $q_{\text{eq}}$ is defined similarly. 
    
\end{itemize}


\bibliographystyle{IEEEtran}
\bibliography{references}

\begin{thebibliography}{10}
\providecommand{\url}[1]{#1}
\csname url@samestyle\endcsname
\providecommand{\newblock}{\relax}
\providecommand{\bibinfo}[2]{#2}
\providecommand{\BIBentrySTDinterwordspacing}{\spaceskip=0pt\relax}
\providecommand{\BIBentryALTinterwordstretchfactor}{4}
\providecommand{\BIBentryALTinterwordspacing}{\spaceskip=\fontdimen2\font plus
\BIBentryALTinterwordstretchfactor\fontdimen3\font minus
  \fontdimen4\font\relax}
\providecommand{\BIBforeignlanguage}[2]{{%
\expandafter\ifx\csname l@#1\endcsname\relax
\typeout{** WARNING: IEEEtran.bst: No hyphenation pattern has been}%
\typeout{** loaded for the language `#1'. Using the pattern for}%
\typeout{** the default language instead.}%
\else
\language=\csname l@#1\endcsname
\fi
#2}}
\providecommand{\BIBdecl}{\relax}
\BIBdecl

\bibitem{al-shawabka_exposing_2020}
A.~{Al-Shawabka}, F.~Restuccia, S.~D'Oro, T.~Jian, B.~C. Rendon, N.~Soltani,
  J.~Dy, K.~Chowdhury, S.~Ioannidis, and T.~Melodia, ``Exposing the
  {{Fingerprint}}: Dissecting the {{Impact}} of the {{Wireless Channel}} on
  {{Radio Fingerprinting}},'' \emph{Proc. of IEEE Conference on Computer
  Communications (INFOCOM)}, p.~10, 2020.

\bibitem{merchant_toward_2019}
K.~Merchant and B.~Nousain, ``Toward {{Receiver}}-{{Agnostic RF Fingerprint
  Verification}},'' in \emph{2019 {{IEEE Globecom Workshops}} ({{GC Wkshps}})},
  Dec. 2019, pp. 1--6.

\bibitem{shen_radio_2021}
G.~Shen, J.~Zhang, A.~Marshall, L.~Peng, and X.~Wang, ``Radio {{Frequency
  Fingerprint Identification}} for {{LoRa Using Deep Learning}},'' \emph{IEEE
  Journal on Selected Areas in Communications}, vol.~39, no.~8, pp. 2604--2616,
  Aug. 2021.

\bibitem{al-shawabka_deeplora_2021}
A.~{Al-Shawabka}, P.~Pietraski, S.~B. Pattar, F.~Restuccia, and T.~Melodia,
  ``{{DeepLoRa}}: Fingerprinting {{LoRa Devices}} at {{Scale Through Deep
  Learning}} and {{Data Augmentation}},'' in \emph{Proceedings of the
  {{Twenty}}-Second {{International Symposium}} on {{Theory}}, {{Algorithmic
  Foundations}}, and {{Protocol Design}} for {{Mobile Networks}} and {{Mobile
  Computing}}}, ser. {{MobiHoc}} '21.\hskip 1em plus 0.5em minus 0.4em\relax
  {New York, NY, USA}: {Association for Computing Machinery}, Jul. 2021, pp.
  251--260.

\bibitem{shen_towards_2021}
G.~Shen, J.~Zhang, A.~Marshall, and J.~Cavallaro, ``Towards {{Scalable}} and
  {{Channel}}-{{Robust Radio Frequency Fingerprint Identification}} for
  {{LoRa}},'' \emph{arXiv:2107.02867 [eess]}, Jul. 2021.

\bibitem{yu_robust_2019}
J.~Yu, A.~Hu, G.~Li, and L.~Peng, ``A {{Robust RF Fingerprinting Approach Using
  Multisampling Convolutional Neural Network}},'' \emph{IEEE Internet of Things
  Journal}, vol.~6, no.~4, pp. 6786--6799, Aug. 2019.

\bibitem{elmaghbub_lora_2021}
A.~Elmaghbub and B.~Hamdaoui, ``{{LoRa Device Fingerprinting}} in the {{Wild}}:
  Disclosing {{RF Data}}-{{Driven Fingerprint Sensitivity}} to {{Deployment
  Variability}},'' \emph{IEEE Access}, vol.~9, pp. 142\,893--142\,909, 2021.

\bibitem{reus-muns_trust_2020}
G.~{Reus-Muns}, D.~Jaisinghani, K.~S. Sankhe, and K.~Chowdhury, ``Trust in {{5G
  Open RANs}} through {{Machine Learning}}: {{RF Fingerprinting}} on the
  {{POWDER PAWR Platform}},'' in \emph{{{IEEE Global Communications
  Conference}}}, 2020.

\bibitem{darpa_radio_nodate}
{DARPA}, ``Radio {{Frequency Machine Learning Systems}},''
  https://www.darpa.mil/program/radio-frequency-machine-learning-systems.

\bibitem{jian_deep_2020}
T.~Jian, B.~C. Rendon, E.~Ojuba, N.~Soltani, Z.~Wang, K.~Sankhe, A.~Gritsenko,
  J.~Dy, K.~Chowdhury, and S.~Ioannidis, ``Deep {{Learning}} for {{RF
  Fingerprinting}}: A {{Massive Experimental Study}},'' \emph{IEEE Internet of
  Things Magazine}, vol.~3, no.~1, pp. 50--57, Mar. 2020.

\bibitem{robinson_novel_2021}
J.~Robinson and S.~Kuzdeba, ``Novel device detection using {{RF}}
  fingerprints,'' in \emph{2021 {{IEEE}} 11th {{Annual Computing}} and
  {{Communication Workshop}} and {{Conference}} ({{CCWC}})}, Jan. 2021, pp.
  0648--0654.

\bibitem{hanna_openset_2020}
S.~Hanna, S.~Karunaratne, and D.~Cabric, ``Open {{Set Wireless Transmitter
  Authorization}}: Deep {{Learning Approaches}} and {{Dataset
  Considerations}},'' \emph{IEEE Transactions on Cognitive Communications and
  Networking}, pp. 1--1, 2020.

\bibitem{orbit_2005}
D.~Raychaudhuri, I.~Seskar, M.~Ott, S.~Ganu, K.~Ramachandran, H.~Kremo,
  R.~Siracusa, H.~Liu, and M.~Singh, ``Overview of the {{ORBIT}} radio grid
  testbed for evaluation of next-generation wireless network protocols,'' in
  \emph{Wireless {{Communications}} and {{Networking Conference}}, 2005
  {{IEEE}}}, vol.~3.\hskip 1em plus 0.5em minus 0.4em\relax {IEEE}, 2005, pp.
  1664--1669.

\bibitem{perahia_next_2013}
E.~Perahia and R.~Stacey, \emph{Next {{Generation Wireless LANs}}: 802.11n and
  802.11ac}.\hskip 1em plus 0.5em minus 0.4em\relax {Cambridge University
  Press}, May 2013.

\bibitem{karunaratne_open_2021}
S.~Karunaratne, S.~Hanna, and D.~Cabric, ``Open {{Set RF Fingerprinting}} using
  {{Generative Outlier Augmentation}},'' in \emph{{{GLOBECOM}} 2021 - 2021
  {{IEEE Global Communications Conference}}}, Dec. 2021.

\bibitem{soltani_more_2020}
N.~Soltani, K.~Sankhe, J.~Dy, S.~Ioannidis, and K.~Chowdhury, ``More {{Is
  Better}}: Data {{Augmentation}} for {{Channel}}-{{Resilient RF
  Fingerprinting}},'' \emph{IEEE Communications Magazine}, vol.~58, no.~10, pp.
  66--72, Oct. 2020.

\bibitem{brown_charrnets_2021}
C.~N. Brown, E.~Mattei, and A.~Draganov, ``{{ChaRRNets}}: Channel {{Robust
  Representation Networks}} for {{RF Fingerprinting}},'' \emph{arXiv:2105.03568
  [cs, eess]}, May 2021.

\bibitem{oyedare_estimating_2019}
T.~Oyedare and J.~J. Park, ``Estimating the {{Required Training Dataset Size}}
  for {{Transmitter Classification Using Deep Learning}},'' in \emph{2019
  {{IEEE International Symposium}} on {{Dynamic Spectrum Access Networks}}
  ({{DySPAN}})}, Nov. 2019, pp. 1--10.

\bibitem{karunaratne_penetrating_2021}
S.~Karunaratne, E.~Krijestorac, and D.~Cabric, ``Penetrating {{RF
  Fingerprinting}}-based {{Authentication}} with a {{Generative Adversarial
  Attack}},'' in \emph{{{ICC}} 2021 - {{IEEE International Conference}} on
  {{Communications}}}, Jun. 2021, pp. 1--6.

\bibitem{karunaratne_real-time_2021}
S.~Karunaratne, S.~Hanna, and D.~Cabric, ``Real-time {{Wireless Transmitter
  Authorization}}: Adapting to {{Dynamic Authorized Sets}} with {{Information
  Retrieval}},'' in \emph{2021 {{IEEE International Symposium}} on {{Dynamic
  Spectrum Access Networks}} ({{DySPAN}})}.\hskip 1em plus 0.5em minus
  0.4em\relax {IEEE}, Dec. 2021.

\end{thebibliography}

	\end{document}